\documentclass[reprint,twocolumn,amsmath,amssymb,prb,superscriptaddress]{revtex4}

\usepackage[pdftex]{graphicx}
\usepackage{dcolumn}% Align table columns on decimal point
\usepackage{bm}% bold math
\usepackage{amsmath}
\usepackage{amssymb}

\newcommand{\bmat}[1]{\bm{\mathrm{ #1}}}
\newcommand{\EQ}[1]{Eq.~(\ref{#1})}

\begin{document}

\title{Efficient multiple time scale molecular dynamics: using colored noise thermostats to stabilize resonances}

\author{Joseph A. Morrone}
\thanks{These authors contributed equally to this work}
\author{Thomas E. Markland}
\thanks{These authors contributed equally to this work}
\affiliation{Department of Chemistry, Columbia University, New York, New York, 10027}

\author{Michele Ceriotti}
\affiliation{Computational Science, Department of Chemistry and Applied Biosciences, ETH Z\"urich, USI Campus, Via Giuseppe Buffi 13, CH-6900 Lugano, Switzerland}

\author{B. J. Berne}
\email{bb8@columbia.edu}
\affiliation{Department of Chemistry, Columbia University, New York, New York, 10027}

\date{\today}

\begin{abstract} Multiple time scale molecular dynamics enhances
computational efficiency by updating slow motions less frequently than
fast motions. However, in practice the largest outer time step
possible is limited not by the physical forces but by resonances
between the fast and slow modes. In this paper we show that this
problem can be alleviated by using a simple colored noise
thermostatting scheme which selectively targets the high frequency
modes in the system. For two sample problems, flexible water and
solvated alanine dipeptide, we demonstrate that this allows the use of
large outer time steps while still obtaining accurate sampling and
minimizing the perturbation of the dynamics.  Furthermore, this
approach is shown to be comparable to constraining fast motions, thus
providing an alternative to molecular dynamics with constraints.
\end{abstract}

\maketitle

\section{Introduction} \label{sec:intro} Over the past two decades
atomistic simulation of chemical, material and biological systems has
become a routine and important tool for understanding, analyzing and
predicting experiment. Time scales of interest now often span into the
microsecond range so as to monitor processes such as protein folding
and transport through membranes. However, the shortest atomic time
scales in these systems, typically for bonded interactions such as
stretches and bends, are of the scale of femtoseconds and therefore
time steps of this order are required to stably evolve the system. As
a result, billions of steps are needed to reach the time scales of
interest, making such simulations a formidable challenge.

Chemical systems typically involve interactions occurring on many time
scales ranging from rapidly varying, but cheap to calculate, bonded
interactions to slow, but expensive, long range
electrostatics. Multiple time scale methods can be used to exploit
this separation in time scales by updating the slow interactions less
frequently than the fast interactions, in principle allowing
significant computational savings to be
achieved.~\cite{streett78,tuckerman90,grubmuller91,tuckerman92}
However, the maximum outer time step which can be obtained is in
practice limited by the resonance between the slow and fast
modes.~\cite{skeel93,ma03_2} This results in energy building up in the
high frequency modes, raising the temperature of the system and
leading to unstable trajectories and incorrect sampling.  As a result
the fastest mode in the system still dictates how frequently the
slowest interactions must be calculated, thereby limiting the maximum
obtainable speed-up.~\cite{izaguirre99,zhou01}

A commonly used approach to delay the resonance barrier is to remove
the high frequencies in the system by constraining
them.~\cite{izaguirre99,zhou01,han07} For example in water one can
constrain the oxygen-hydrogen bond length and intramolecular
hydrogen-hydrogen distance. This allows a larger outer time step to be
used but requires prior knowledge of the coordinates that comprise the
fast modes. In simple systems these coordinates may be known, but in
complex biological and materials systems this is not always the case
and one could risk constraining a degree of freedom which has vital
importance to mechanism or function. The MOLLY method can be viewed in
a similar fashion since it requires prior specification of coordinates
of fast modes which are then used to filter out the destabilizing
components of the slow forces.~\cite{izaguirre99,izaguirre01,ma2003}

It is now well established that large time steps can be achieved while
retaining full flexibility by coupling the system to a
bath.~\cite{barth98,qian02,minary04} We note that the bath serves two
purposes: to remove energy which builds up in the modes and to disrupt
the high frequency modes from resonating with the lower
frequencies. The most common choice is to couple each atom in the
system to a white noise Langevin (WNL) bath which acts uniformly
across the spectrum of the system.~\cite{barth98,qian02} However, as
has been pointed out previously,~\cite{izaguirre01} the strength of
the system-bath coupling needed to stabilize large time steps
significantly disrupts the motion of the slow modes thereby greatly
hindering diffusive and orientational motion. As we will show this
leads to a situation where any computational speed-ups gained by
increasing the outer time step are largely outweighed by the decrease
in the rate at which different configurations are explored. To avoid
this one would ideally like to couple strongly to high frequency
motion in the system while leaving low frequencies unperturbed. This
can be achieved by transforming to the normal modes of the system at
each time step and then evolving using a strong coupling to high
frequencies with weak coupling to low ones. Indeed, it has been shown
for a simple biological system in implicit solvent that this approach
is very successful at removing resonance issues while preserving
dynamics.~\cite{sweet08} However, in practice performing normal mode
transformations at each step for large systems is prohibitively
expensive.

In this paper we attempt to reconcile the simplicity of the WNL
approach with the effectiveness of the targeted normal mode approach
by using tailored colored noise. Recent work has made this possible by
showing how the generalized Langevin equation (GLE) can be used to
thermostat molecular dynamics simulations using an extended WNL
formalism.~\cite{ceriotti_09_1,ceriotti_09_2,ceriotti_10_1,ceriotti_10_2,ceriotti_10_3}
Unlike white noise, colored noise can be tailored to have a frequency
dependent coupling which allows for much greater flexibility in its
application. For example a simple colored noise thermostat was
designed for use in Car-Parrinello {\it ab-initio} molecular dynamics
simulations which only targets atomic motion while allowing the high
frequency fictitious electronic degrees of freedom to evolve
freely.~\cite{ceriotti_09_1} In this work we will demonstrate how
colored noise can instead be used to design a bath which heavily damps
high frequencies while leaving low frequencies largely unaffected. The
simple scheme which results allows the resonance barrier to be
postponed facilitating the use of large outer time steps while
yielding accurate sampling and minimal impact on the
dynamics. Applications of this approach to flexible water and an
aqueous solution of alanine dipeptide demonstrate that significant
increases in computational efficiency can be achieved while requiring
little \emph{a priori} knowledge of the system.

The outline of the paper is as follows. Section II briefly reviews the
colored noise thermostatting approach and shows how colored noise
profiles can be constructed in a transparent way by combining simple
analytic forms. The implementation of the GLE thermostat in a standard
reference system propagator algorithm (RESPA)~\cite{tuckerman92} is
then discussed. Section III outlines the simulations performed using
this scheme to calculate the static and dynamic properties of a fully
flexible model of liquid water and an aqueous solution of alanine
dipeptide. Section IV discusses the results of these applications and
Sec. V concludes.

\section{Theory} \label{sec:theory}
\subsection{Colored noise thermostats} For a particle of mass $m$ with
position $x$ and momentum $p$ moving on a one dimensional potential
energy surface $V(x)$, the generalized Langevin equation
is,~\cite{berne77,zwanzig,gardiner}
\begin{eqnarray} \dot{x} &=& p/m \\ \dot{p} &=& f(x) - \int\limits_0^t
\mathrm{d}\tau K(t-\tau) p(\tau) + R(t) \label{eq:genlang}
\end{eqnarray} where $K(t)$ is the memory kernel, $R(t)$ is a
non-Markovian ``colored'' random force and $f(x) = -\mathrm{d}V(x) /
\mathrm{d}x$ is the force due the potential. The
fluctuation-dissipation theorem dictates that for an equilibrium
system at temperature $T$, $K(t)$ and $R(t)$ are related by,
\begin{eqnarray} m k_B T \; K(t) &=& \left< R(t) R(0) \right>.
\label{eq:fluc_dis}
\end{eqnarray} where $k_B$ is the Boltzmann constant.

In the case where the random force is uncorrelated, $\left< R(t) R(0)
\right> = b^{2}\delta(t)$, the white noise Langevin equation,
\begin{eqnarray} \dot{x} &=& p/m \\ \dot{p} &=& f(x) - \gamma p + b \;
\xi(t) \label{eq:wnllang}
\end{eqnarray} is recovered. Here $\xi(t)$ is a Gaussian Markov
process with unit variance and the fluctuation dissipation theorem in
\EQ{eq:fluc_dis} reduces to,
\begin{eqnarray} 2 m k_B T \gamma = b^2
\end{eqnarray} which is commonly used as a tool to thermostat
molecular dynamics simulations.

A colored noise thermostat can be implemented by exactly mapping
Equation \ref{eq:genlang} onto a Markovian dynamics in an extended
space consisting of a set of auxiliary momenta, $\bm{s}$, that are
coupled to the system momentum, $p$, in the presence of a Markovian
bath.~\cite{ferrario,marchesoni,ceriotti_09_1,ceriotti_09_2,ceriotti_10_1,ceriotti_10_2}
The equations of motion are given by:
\begin{eqnarray} \dot{x} &=& p/m \\ \left( \begin{array}{c} \dot{p} \\
\dot{\bm{s}} \end{array} \right) &=& \left( \begin{array}{c} f(x) \\
\bm{0} \end{array} \right) - \bmat{\Gamma} \left( \begin{array}{c} {p}
\\ {\bm{s}} \end{array} \right) + \bmat{B}
\bm{\xi}(t) \label{eq:langmat}
\end{eqnarray} where $\bm{\xi}(t)$ is a vector of uncorrelated
Gaussian noise.  The drift (or damping) matrix ($\bmat{\Gamma}$), may
be related to the diffusion matrix ($\bmat{B}$) by a recasted
fluctuation-dissipation theorem.~\cite{gardiner,ceriotti_10_1}
\begin{eqnarray} m k_{B}T \left( \bmat{\Gamma} +
\bmat{\Gamma}^\text{T} \right) &=& \bmat{B} \bmat{B}^\text{T}.
\end{eqnarray} The matrix elements of $\bmat{\Gamma}$,
\begin{eqnarray} \bmat{\Gamma} &=& \left( \begin{array}{cc}
\gamma_{pp} & \bm{\gamma}_{ps}^\text{T} \\ \bm{\gamma}_{sp} &
\bmat{\Gamma}_{ss} \end{array} \right) \label{eq:acomp}
\end{eqnarray} determine the form of the memory kernel that acts on
the system via
\begin{eqnarray} K(t) &=& 2 \gamma_{pp} \delta(t) -
\bm{\gamma}_{ps}^\text{T} \; e^{-|t| \bmat{\Gamma}_{ss}} \;
\bm{\gamma}_{sp} \label{eq:memory}.
\end{eqnarray} If the extended momenta are uncoupled from the system
equation, \EQ{eq:langmat} reduces to the white noise Langevin equation
(\EQ{eq:wnllang}) with $\gamma_{pp}=\gamma$. In the next section we
will describe how \EQ{eq:memory} can be used to design a drift matrix
that yields a colored noise profile such that the high frequency modes
are strongly coupled and slow modes weakly coupled to the bath.

\subsection{Designing the colored noise profile} The effect of the
bath on the system is determined by the details of $K(t)$.  Using the
extended white noise Langevin framework outlined above the effect of
the bath and the form of the memory kernel is determined by
customizing the drift matrix $\bmat{\Gamma}$ (see \EQ{eq:memory}).

The expression given in the right hand side of \EQ{eq:memory} consists
of two terms, a simple white noise component of strength $\gamma_{pp}$
and one that depends upon the coupling of the auxiliary momenta.  This
second term is a scalar and is therefore invariant to the choice of
basis. The time dependence of the second term of \EQ{eq:memory} is
contained in the matrix exponential of submatrix, $\bmat{\Gamma}_{ss}$
and it is therefore transparently expressed in its eigenbasis.  The
eigendecomposition of the submatrix yields $\bmat{\Gamma}_{ss} =
\bmat{U}\bmat{\Lambda} \bmat{U}^{-1}$, where
$\bmat{\Lambda}_{ij}=\lambda_j \delta_{ij}$ is the diagonal eigenvalue
matrix. Matrix $\bmat{U}$ is comprised of the set of eigenvectors in
its columns and may be used to rotate, $\bmat{\Gamma}_{ss}$ into the
eigenbasis. The memory function is now expressed as,
\begin{eqnarray} K(t) &=& 2 \gamma_{pp} \delta(t) - \bm{\gamma}_{ps}^T
\bmat{U} e^{-|t| \bmat{\Lambda} } \bmat{U}^{-1}
\bm{\gamma}_{sp}. \label{eq:memeigen}
\end{eqnarray} In order to recover correct equilibrium behavior in the
$t \rightarrow \infty$ limit, it is a necessary but not sufficient
condition that the real part of the eigenvalues of
$\bmat{\Gamma}_{ss}$ are chosen to be
positive.~\cite{gardiner,ceriotti_10_1} The resultant kernel consists
of a linear combination of exponential decaying forms, arising from
the real eigenvalues, and damped oscillatory ones from the complex
conjugate pairs.

In considering the effect of a colored noise bath it is physically
intuitive to consider the memory spectrum, $\hat{K}(\omega)$, which is
the Fourier transform of the memory kernel.  Given that the memory
function may be expressed as a linear combination of a white noise
term and a set of exponential forms with real or complex conjugate
pairs of eigenvalues, the memory spectrum is readily computable.  For
a given eigenvalue $\lambda_j= a_j+ib_j$, the Fourier transform of the
corresponding exponential function is a Lorentzian, \begin{eqnarray}
\int \limits_{-\infty}^{\infty} \; \mathrm{d}t \; e^{-i \omega t}
e^{ib_j t} e^{-a_j |t|} &=& 2\; \frac{a_j}{a_j^2+(\omega - b_j)^2}
\end{eqnarray} The width and center of the Lorentzian is given by the
real and imaginary parts of the corresponding eigenvalue,
respectively.  Therefore, the memory spectrum is expressible as a
constant $2\gamma_{pp}$ from which a set of Lorentzian functions are
added or subtracted.  In order to generate the correct equilibrium
distribution, forms must be chosen such that $\hat{K}(\omega)$ is
always greater than zero.~\cite{berne71,ceriotti_10_1}

Although the form of the memory function as given by \EQ{eq:memeigen}
appears simple, ``translating'' a chosen form into a stable drift
matrix is non-trivial.  This is due to the fact that although the
basis in which the submatrix $\bmat{\Gamma}_{ss}$ is expressed is
arbitrary with respect to the form of the memory function, this choice
of basis is crucial for the generation of stable dynamics within the
colored noise thermostat implementation.  Namely, a requirement of the
integration of the equations of motion is that the symmetric part of
$\bmat{\Gamma}$ must be positive definite (see Sec. \ref{ssec:int}).
Indeed, it is possible to choose forms of the kernel based on
\EQ{eq:memeigen} such that $\hat{K}(\omega)>0$ while violating this
condition.

Despite these difficulties, it is possible to construct drift matrices
which correspond to a memory spectrum of a desired form where the
matrix elements are transparently relatable to the the time
dependence, namely the eigenvalues of $\bmat{\Gamma}_{ss}$ are
explicit parameters of $\bmat{\Gamma}$.  Drift matrices of this sort
have already appeared in the
literature,~\cite{ceriotti_09_1,ceriotti_10_2} and their details are
summarized in Appendix \ref{app:forms}.  We now present such a form
that is tailored for the problem at hand.

\begin{figure}
\begin{center}
\includegraphics[scale=0.33]{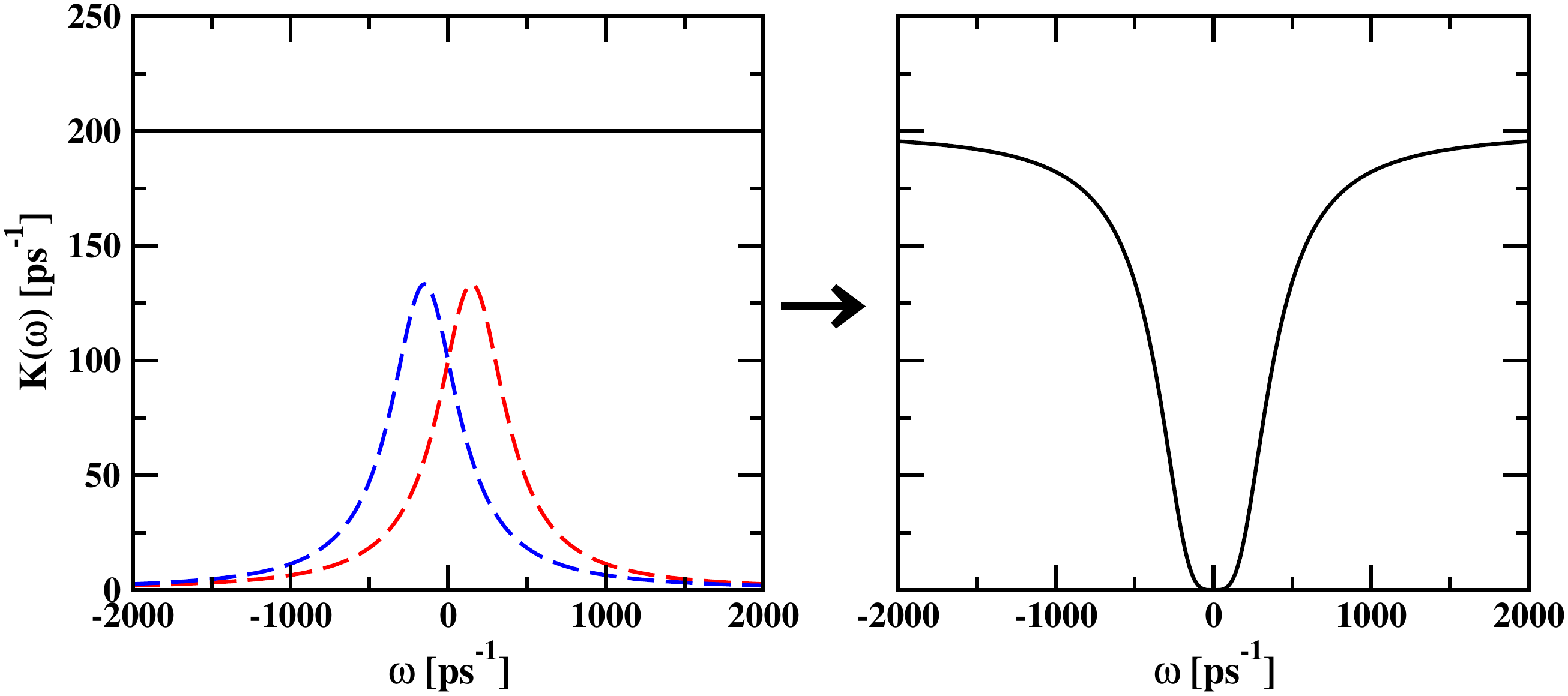}
\caption{The shape of the memory spectra given by \EQ{eq:3x3spec} is
detailed schematically.  In the left panel the white noise component
(black line) is shown alongside Lorentzians centered at $\pm
\tilde{\omega}/2$ (blue and red dashed curves).  The right panel shows
the memory spectra which results from the subtraction of the sum of
the Lorentzians from the white noise component.  The spectra is small
near zero and plateaus to the value of the white noise component at
large $\omega$.  } \label{fig:3x3}
\end{center}
\end{figure}

In selecting the details of the colored noise profile it is
instructive to first consider the memory spectrum, which provides
information about how strongly the modes of the system exchange energy
with the bath at a given frequency.~\cite{berne71} Additionally, the
$\omega=0$ value of the memory spectrum is inversely proportional to
the diffusion constant of a free particle ($f(x)=0$) coupled to the
bath.  Therefore, if a profile is required that couples the bath
strongly to high frequency modes and more weakly to low ones, one can
start by constructing a memory kernel whose spectrum is small at low
frequencies and large near the frequencies where strong coupling is
desired.  The following drift matrix,
\begin{widetext}
\begin{eqnarray} \bmat{\Gamma} &=& \left( \begin{array}{ccc}
\gamma_\infty & 3^{1/4} \sqrt{ \tilde{\omega}(\gamma_\infty-\gamma_0)
} & \frac{1}{3^{1/4}} \sqrt{ \tilde{\omega}(\gamma_\infty-\gamma_0) }
\\ 3^{1/4} \sqrt{ \tilde{\omega}(\gamma_\infty-\gamma_0)} &
\tilde{\omega}\sqrt{3} & \tilde{\omega} \\ -\frac{1}{3^{1/4}} \sqrt{
\tilde{\omega}(\gamma_\infty-\gamma_0) } & -\tilde{\omega} &
0 \end{array} \right) \label{eq:3x3}
\end{eqnarray} corresponds to a memory spectrum which possesses such a
form.  The associated memory kernel and spectrum are given by the
following equations:
\begin{eqnarray} K(t) &=& 2 \gamma_\infty \delta(t) - 2\;
\frac{\tilde{\omega}}{\sqrt{3}} (\gamma_\infty-\gamma_0)
e^{-\tilde{\omega}\sqrt{3}|t|/2} \cos(\tilde{\omega}
t/2) \label{eq:3x3mem} \\ \hat{K}(\omega) &=& 2 \gamma_\infty - 2\;
\frac{\tilde{\omega}}{\sqrt{3}} (\gamma_\infty-\gamma_0) \left( \frac{
\frac{\sqrt{3}}{2} \tilde{\omega}}{ \frac{3}{4} \tilde{\omega}^{2} +
(\omega-\frac{\tilde{\omega}}{2})^2} + \frac{ \frac{\sqrt{3}}{2}
\tilde{\omega}}{\frac{3}{4}\tilde{\omega}^{2} +
(\omega+\frac{\tilde{\omega}}{2})^2} \right).  \label{eq:3x3spec}
\end{eqnarray} 
\end{widetext}
The memory spectrum of \EQ{eq:3x3spec} is shown
schematically in Figure \ref{fig:3x3}. It can be understood as a white
noise of strength $\gamma_\infty$ from which a pair of Lorentzians of
width $\tilde{\omega} \sqrt{3}/2$ centered at $\omega=\pm
\tilde{\omega}/2$ is subtracted.  In this way the Lortenzian terms can
be thought of as partially ``canceling out'' the white noise component
at low frequencies.  The shape is such that its value is $2\gamma_0$
at $\omega=0$ and approaches a value $2 \gamma_\infty$ as $\omega
\rightarrow \infty$.  In order to fulfill our design requirements of
the noise profile, we take the parameter $\gamma_0$ to be an
arbitrarily small (non-zero) number.  The parameter $\tilde{\omega}$
is related to the frequency at which the memory spectrum begins to
rise rapidly.  When the parameters are chosen to have positive values,
$\hat{K}(\omega)$ is an even, positive definite function, as is
required to generate the correct equilibrium distribution.

The colored noise profile expressed in Eqs.  (\ref{eq:3x3}),
(\ref{eq:3x3mem}), and (\ref{eq:3x3spec}) provides a transparent and
readily tunable form for the drift matrix of the colored noise
thermostat with an associated memory spectrum of the shape that we
desire.  However the coupling of the bath to system is non-local, and
although intuition can inform our choice of the shape of the memory
spectrum, one cannot \emph{a priori} determine the exact values of the
parameters $\gamma_\infty$ and $\tilde{\omega}$ which adequately
couple to high frequency modes and simultaneously minimize the
perturbation on low frequency modes.

In previous work, colored noise thermostats have been successfully
tuned according to the energy relaxation times of harmonic oscillators
coupled to the thermostat.~\cite{ceriotti_10_1} However in this work
we find it more appropriate to estimate the disturbance of the
spectrum of a test harmonic oscillator engendered by the colored noise
thermostat.  The velocity autocorrelation function (and hence
spectrum) for an oscillator with frequency $\omega_0$ in the presence
of the colored noise thermostat may be exactly computed by matrix
algebra.~\cite{tuckerman93,ceriotti_10_1} The spectrum is both
broadened and shifted with respect to the free oscillator result, and
the size of these differences provides some measure of the strength of
the system-bath coupling. For frequencies that are small compared to
the bath parameter $\tilde{\omega}$ and when $\gamma_0$ is negligible,
the ratio of the peak position of the system-bath spectrum
($\omega_P$) to that of the free spectrum ($\omega_0$) may be
estimated by the following expression (see Appendix \ref{app:shift}),
\begin{eqnarray} \frac{\omega_P}{\omega_0} &=&\frac{1}{\sqrt{1+
\frac{\gamma_\infty}{\tilde{\omega}\sqrt{3}}}} + \mathcal{O}\left(
\left( \frac{\omega_0}{\tilde{\omega}}\right)^2
\right) \label{eq:shift}.
\end{eqnarray} It is important to note that this estimate depends on
the value of the friction in the high frequency limit, and is a
function of the ratio $\gamma_\infty / \tilde{\omega}$.  This
underscores the non-locality in frequency space of the system-bath
interaction.

Equation \ref{eq:shift} provides an estimate of the impact of the
colored noise thermostat on the low frequency modes.  In order to
ensure that the high frequency modes are sufficiently damped, the
parameter $\gamma_\infty$ is set to be large. A lower bound is
provided by the white noise friction that is required to stabilize any
resonance instabilities that are present in the system.  Fine tuning
can then be accomplished by testing the performance of a set of noise
profiles on a realistic system in conjunction with a multiple time
scale integrator.  However, since most biomolecular systems have
similar spectral features the profiles presented here should provide
good performance in a broad range of studies without
reparameterization.  The parameters for the colored noise profiles
that will be utilized in this study are given in Table \ref{tab:tab1}.

\begin{figure}
\begin{center}
\includegraphics[scale=0.33]{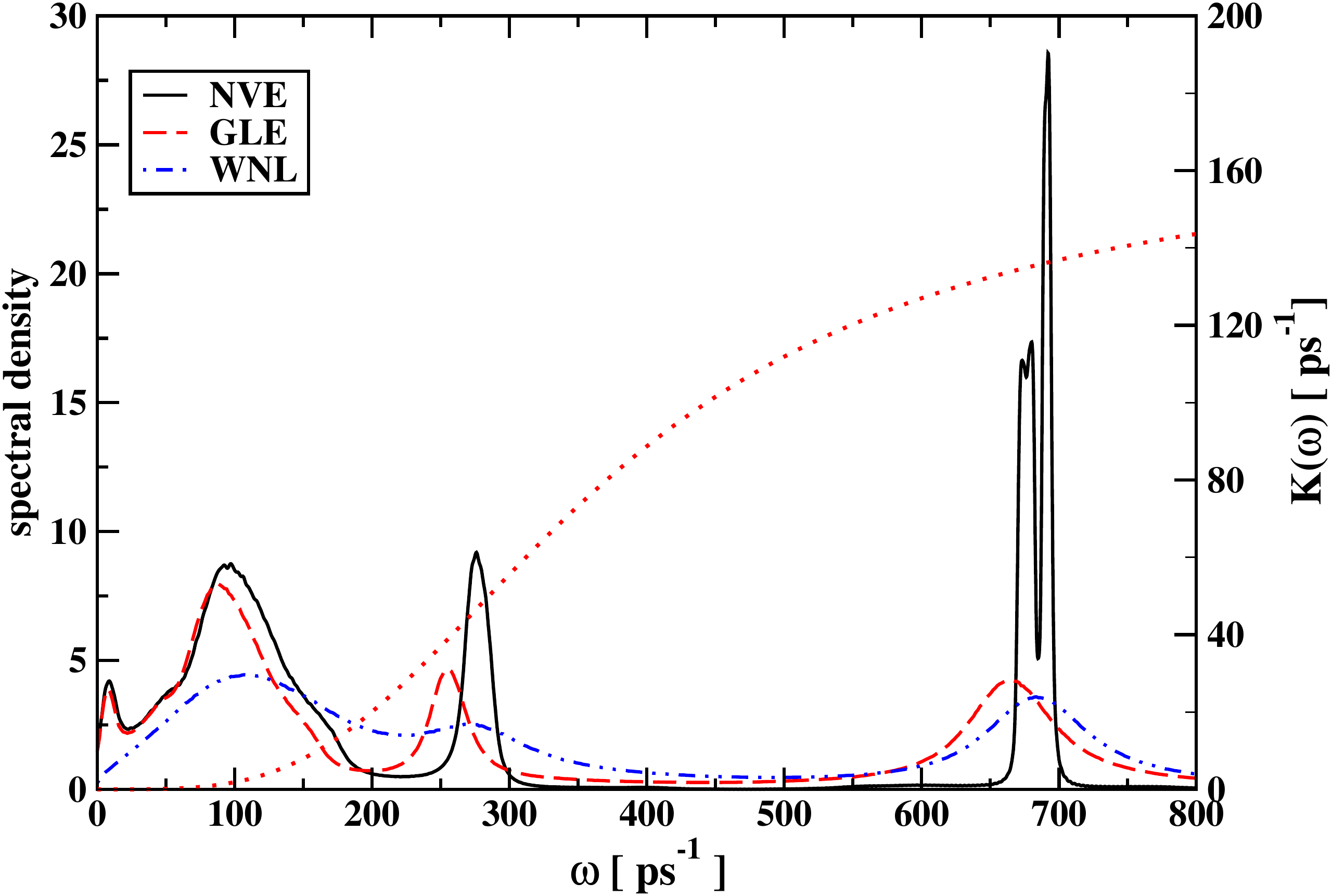}
\caption{The spectra of SPC/Fw water computed from a microcanonical
simulation (solid, black line), and simulations that employ the GLE -
12 fs colored noise parameter set (red, dashed line), and a white
noise (blue, dot-dashed line) Langevin thermostat.  The memory spectra
that is obtained from the GLE dynamics is given by the red, dotted
line. The strength of the white noise bath corresponds to the $\omega
\rightarrow \infty$ limit of this profile ($\gamma=83.3$
ps$^{-1}$). } \label{fig:spec}
\end{center}
\end{figure}

In order to gauge the impact of the colored noise thermostat, the
spectrum of a flexible water model~\cite{wu06} in the presence of a
bath as defined by parameter set GLE - 12 fs in Table \ref{tab:tab1}
is plotted in Figure \ref{fig:spec}.  The result is compared to
microcanonical and white noise Langevin dynamics with a friction that
is equivalent to the $\omega \rightarrow \infty$ limit of the colored
profile.  It can be readily seen that, when compared with the
microcanonical dynamics, the lower frequency modes are less disturbed
than those related to bending and stretching.  It is the
intramolecular modes that have been targeted for damping and one can
see that the impact of the colored bath on the oxygen-hydrogen stretch
at $\approx 680$ ps$^{-1}$ is comparable to that engendered by the
high-friction white noise bath.  Furthermore we note that the ratio of
the low frequency peak positions of the thermostatted spectra to the
microcanonical peak positions is approximately $90\%$.  This value is
close to the estimate provided by \EQ{eq:shift}, which predicts a
shift of $93\%$ (see Table \ref{tab:tab1}).

\begin{table}
\begin{center}
\begin{tabular}{||c|c|c|c|c||} \hline matrix no. & $\gamma_\infty$
(ps$^{-1}$) & $\gamma_0$ (ps$^{-1}$) & $\tilde{\omega}$ (ps$^{-1}$) &
$\omega_P / \omega_0$ \\ \hline GLE - 12fs & 83.33 & 0.01 & 300.0 &
0.93 \\ GLE - 16fs & 125.0 & 0.01 & 100.0 & 0.76 \\ GLE - 20fs & 200.0
& 0.01 & 75.0 & 0.63 \\ \hline
\end{tabular}
\caption{The parameters sets of the drift matrix (\EQ{eq:3x3}) which
are utilized in this study.  Parameter sets are labeled according to
the outer time step which they are used in conjunction with (See
Sec. \ref{sec:details}).  } \label{tab:tab1}
\end{center}
\end{table}

\subsection{Integration scheme} \label{ssec:int}

We now briefly outline the numerical method used to evolve the system
according to the equations of motion in \EQ{eq:langmat} and their
integration within a standard RESPA multiple time scale scheme. For
clarity the following equations are shown for a single degree of
freedom. However, extension to more dimensions follows directly.

Integrators for molecular dynamics can be generated Trotter
factorization of the Liouville propagator.~\cite{tuckerman92} A
suitable factorization to evolve the equations of motion of a system
coupled to a colored noise thermostat (\EQ{eq:langmat}) over a time
step $\Delta t$ is given
by,~\cite{bussi07,ceriotti_10_1,ceriotti_10_3}
\begin{equation} e^{i L \Delta t} = e^{i L_{ps}\Delta t/2} e^{i L_{p}
\Delta t/2} e^{i L_{x}\Delta t} e^{i L_{p} \Delta t/2} e^{i L_{ps}
\Delta t/2}.
\end{equation} Each factor provides an analytic operation on the state
of the system.  The operator $e^{i L_{x} \Delta t}$ provides a
coordinate shift through a time step $\Delta t$,
\begin{equation} x \leftarrow x + \Delta t \frac{p}{m}
\end{equation} while $e^{i L_{p} \Delta t}$ evolves the momenta by a
time increment $\Delta t$,
\begin{equation} p \leftarrow p + \Delta t f(x).
\label{eq:evolp}
\end{equation} The combination of these two operations is the standard
velocity Verlet algorithm~\cite{tuckerman92}. The outermost operation
$e^{i L_{ps} \Delta t}$ provides the effect of the colored noise
thermostat on the system momentum, $p$ and evolves the additional
thermostat degrees of freedom, $\bm{s}$. This operation can be shown
to be,~\cite{ceriotti_10_1,ceriotti_10_3}
\begin{equation} \bm{p} \leftarrow {\bf C}_1 \; \bm{p} +\sqrt{m k_{b}
T}{\bf C}_2\,\bm{\xi} \label{eq:thermev}
\end{equation} where $\bm{\xi}$ is a vector of independent Gaussian
numbers. Here
\begin{equation} \bm{p} = \begin{pmatrix} p \cr \bm{s}
\cr \end{pmatrix}
\end{equation} is a vector containing the system momentum, $p$, and
extended momenta $\bm{s}$, and
\begin{equation} {\bf C}_1 = e^{-(\Delta t/2) {\bf \Gamma}},
\end{equation} and
\begin{equation} {\bf C}_2^T{\bf C}_2 = {\bf I}-{\bf C}_1^T{\bf
C}_1.  \label{eq:c2}
\end{equation} Again, it is useful to note that this reduces to a
standard integrator for the white noise Langevin
thermostat~\cite{bussi07} when the extended momenta are decoupled from
the system. In order to recover $\bmat{C}_2$ a Cholesky decomposition
must be performed on the expression in \EQ{eq:c2}.  This operation
requires that the symmetric part of the matrix $\bmat{\Gamma}$ is
positive definite.~\cite{ceriotti_10_1}

When the system consists of more than the one degree of freedom 
the ability of the extended momenta to respond to the different
frequencies experienced by each particle requires a local (massive) coupling.~\cite{ceriotti_10_1}
Hence for a system of N particles consisting of 6N variables (positions and
momenta) the colored noise thermostat corresponding to the drift
matrix in \EQ{eq:3x3} adds 6N additional variables in the form of the
auxiliary momenta, $\bm{s}$. This compares favorably with those required in
local Nose Hoover schemes which add 18N variables if a typical
chain of length three is chosen.~\cite{martyna92,martyna96} Evolution of the auxiliary
momenta in \EQ{eq:thermev} is a $3\times 3$ matrix multiplication and 3N of these
operations are required for each thermostat evolution of the N
particle system. The local nature of the thermostat makes the
operation easily parallelizable. For all the systems considered in
this study we found the cost of the thermostat operations to be small
in comparison to the force calculations which dominate the
computational cost.

To construct a multiple time scale scheme the forces are partitioned
into a sum of rapidly and slowly varying components.  In the case of
three components, we may write the total force as,
\begin{eqnarray} f(x) = f^{(1)}(x) + f^{(2)}(x) +
f^{(3)}(x) \label{eq:fsplit}
\end{eqnarray} where $f^{(1)}(x)$ corresponds to the slowest forces,
which can be integrated with the largest time step, and $f^{(3)}(x)$
the fastest ones which necessitate the smallest time step for stable
integration.  With this splitting of the forces the Liouville operator
can be factorized as,
\begin{widetext}
\begin{eqnarray} e^{i L \Delta t} &=& e^{i L_{p}^{(1)} \Delta t/2}
\times \nonumber \\ && \prod^{M_{2}} \left[ e^{i L_{ps} \Delta
t_{2}/2} e^{i L_{p}^{(2)} \Delta t_{2}/2} \prod^{M_{3}} \left( e^{i
L_{p}^{(3)} \Delta t_{3}/2} e^{i L_{x} \Delta t_{3}} e^{i L_{p}^{(3)}
\Delta t_{3}/2} \right) e^{i L_{p}^{(2)} \Delta t_{2}/2} e^{i L_{ps}
\Delta t_{2}/2} \right] \nonumber \\ && \times \; e^{i L_{p}^{(1)}
\Delta t/2}
\end{eqnarray} 
\end{widetext}
where the exponents of $L_{p}^{(k)}$ perform the
evolution shown in Eq. \ref{eq:evolp} under the force
$f^{(k)}(x)$. The time steps for the integration of the medium and
fast forces are then,
\begin{eqnarray} \Delta t_{2} = \Delta t / M_{2}
\end{eqnarray} and
\begin{eqnarray} \Delta t_{3} = \Delta t / (M_{2} M_{3})
\end{eqnarray} respectively, where the whole numbers $M_{2}$ and
$M_{3}$ are chosen to be sufficiently large so as to allow stable
integration of the system under the forces $f^{(2)}(x)$ and
$f^{(3)}(x)$.  The thermostat evolution is located in the middle loop
so that the bath is efficiently coupled to the fast system motions.

\section{Computational Details} \label{sec:details} The colored noise
RESPA scheme introduced in the previous section was used to perform
simulations of pure water and alanine dipeptide in explicit water. The
water system consisted of 1000 molecules simulated in a 31.07
$\text{\AA}$ box. The alanine dipeptide was described by the OPLS-AA
force field~\cite{oplsaa,kaminski01} in a 20 $\text{\AA}$ box
containing 252 water molecules. The SPC/Fw potential~\cite{wu06} was
used to model the interactions between water molecules. All bonded
terms (stretches, bends and torsions) were treated as flexible.

The forces were partitioned into three levels as in
Eq. \ref{eq:fsplit}. The bonded terms were updated every 0.5 fs and
the non bonded interactions below 9 $\text{\AA}$ were updated every 2
fs. The long range electrostatics, which dominate the computational
cost of the simulation, were updated at the outer time step which was
varied between 2 and 20 fs as discussed below. The electrostatics were
partitioned into short and long range parts using the RESPA2
scheme.~\cite{stuart96,zhou01} The function that switches between
regions is a quintic spline which acts over a 4 $\text{\AA}$
region.~\cite{morrone10} The cut off between the short and long range
electrostatics was chosen so as to optimize the computational
performance within our code.  Larger cutoffs have been shown to
produce better stability within the MTS formalism~\cite{han07} so for
maximum benefit the cutoffs used should be balanced with performance
depending on the exact implementation.

Due to the cost of calculating the long range electrostatic
interactions we wish to make the outer time step as large as possible
while still obtaining correct sampling. We therefore performed
simulations using time steps of 12 fs, 16 fs, and 20 fs. Resonance
artifacts are extremely pronounced at these outer time steps so the
simulation must be stabilized with either strong white noise damping
or optimized colored noise thermostats. The baseline results for
dynamic and equilibrium properties were obtained from a microcanonical
and a white noise Langevin simulation, respectively and were performed
using an outer time step of 2 fs.  For the microcanonical results and each 
combination of thermostat
and time step, water statistics were collected over twelve independent runs 
with initial velocities sampled from the Boltzmann distribution. For water a total 
of 12 ns of simulation time was performed, while for alanine dipeptide 400 ns was necessary to
converge the properties reported.

\section{Results and Discussion} \label{sec:results} For systems that
exhibit large resonance artifacts, we now show that an appropriately
designed colored noise thermostat is capable of yielding accurate
sampling while simultaneously minimizing the thermostat's impact on
diffusional and orientational dynamics.  In order to test this scheme,
we perform simulations on flexible water and a fully flexible
simulation of aqueous alanine dipeptide. The broad spectrum of
frequencies in aqueous systems range from fast intramolecular
stretches to slow diffusional modes (see Figure \ref{fig:spec}).  The
strong coupling between the modes makes this a challenging example to
test our approach. Hence, for flexible water simulations the resonance
barrier occurs at an outer time step of $\approx 3$ fs when the
microcanonical r-RESPA algorithm is employed.~\cite{ma03_2}

In this work, we design colored noise thermostats that are capable of
stabilizing resonance artifacts for outer time steps of 12, 16, and 20
fs while ensuring that the error in the energies is within $0.5\%$ of
the baseline results. The parameters for these thermostats are given
in Table \ref{tab:tab1}. In order to provide comparison as to the
effectiveness of our scheme, we also perform simulations that use a
white noise Langevin (WNL) thermostats to yield comparable
accuracy. These runs utilize a friction of $14.2$ ps$^{-1}$, 40.0
ps$^{-1}$, and 100.0 ps$^{-1}$ in conjunction with outer times steps
of 12 fs, 16 fs, and 20 fs, respectively.

Additionally, recent work has shown that when rigid constraints are
placed on the fastest degrees of freedom, a weak Langevin coupling of
$\gamma = 1$ ps$^{-1}$ is required to stabilize the simulation at an
outer time step of 12 fs.~\cite{han07} We therefore perform
simulations using a 2 fs outer time step with this friction.  This
facilitates a comparison of the dynamical perturbation arising from
constrained dynamics using white noise stabilization with that caused
by fully flexible dynamics using our colored noise scheme.

\subsection{Water} \label{ssec:water}

We first apply our scheme to pure flexible water.  In Table
\ref{tab:tab2} the average temperature and average bonded and
non-bonded potential energies given for combinations of outer time
step and method of resonance stabilization, either white noise
Langevin (WNL) or generalized (colored noise) Langevin (GLE)
thermostatting. The baseline result to which the static properties of
all stabilized runs are compared utilizes an outer time step of $2$ fs
and a WNL thermostat with a friction of $\gamma=1$ ps$^{-1}$.  It can
be seen that the GLE runs reproduce baseline results to within an
error of $0.5\%$ for both the bonded and non-bonded components of the
energy.  The chosen white noise Langevin couplings exhibit comparable
overall performance at each corresponding outer time step.  Upon study
of Table \ref{tab:tab2}, it can be seen that the error is up to four
times larger in the bonded energy as compared to the non-bonded energy
for the GLE runs.  In the case of the WNL stabilized simulations, it
is up to ten times larger.  This can be explained by the fact that
resonance instabilities are most severe in the high frequency
intramolecular modes.  Therefore, as the error is not uniformly
distributed across the system, it is of great utility to consider
different components of the energy when assessing resonance
stabilization.  Although the error in the temperature is slightly
elevated for runs with an outer time step of 20 fs, it is within
$0.5\%$ for all other results.  The errors in temperature tend to
correlate with those in the potential energy.  However, from the
discussion above, the potential energy can be seen to be a more
sensitive measure of sampling accuracy.

\begin{figure}
\begin{center}
\includegraphics[scale=0.37]{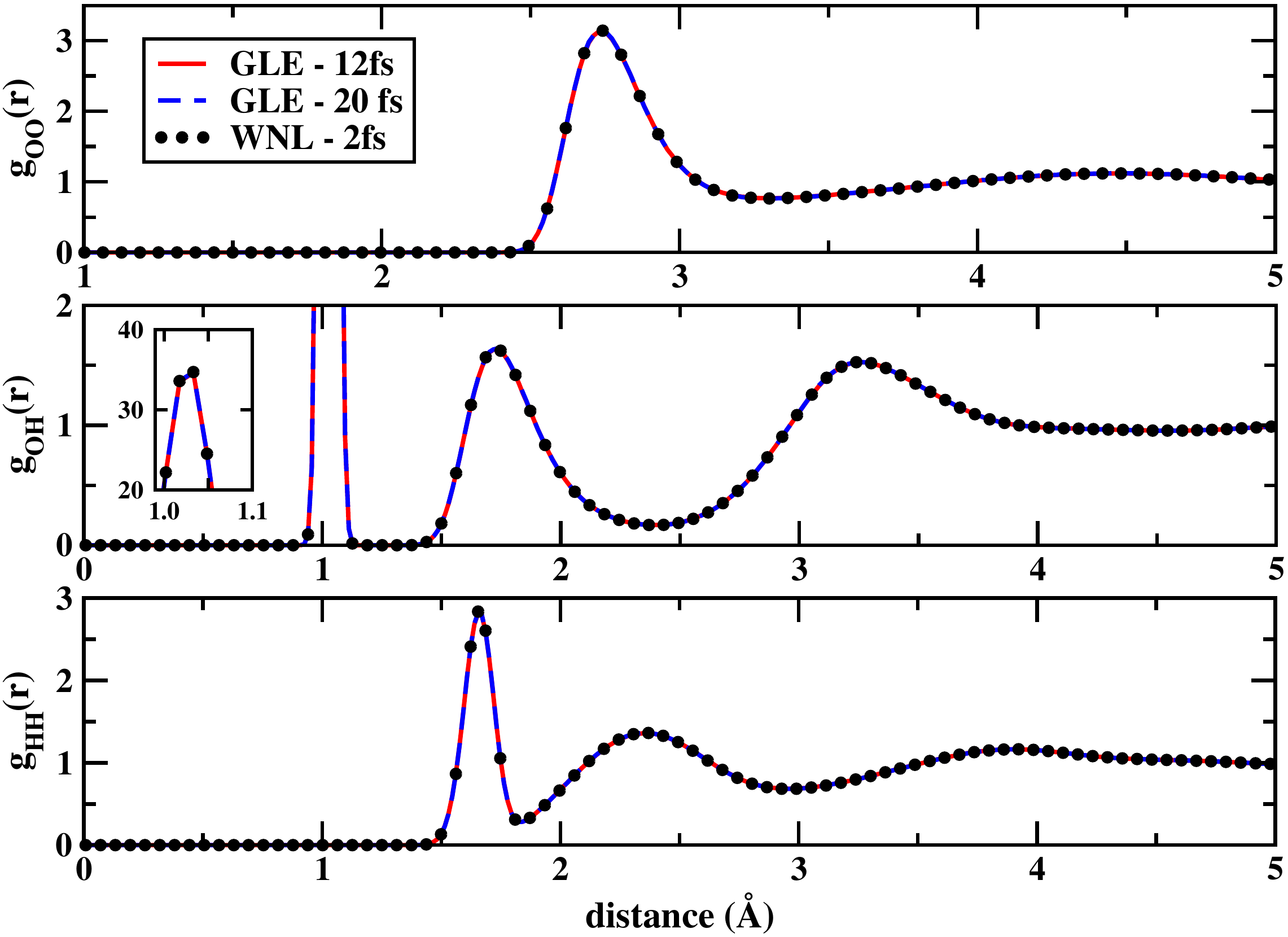}
\caption{The atom-atom radial distribution functions of water computed
from the baseline MTS white noise Langevin simulation with an an outer
time step of 2 fs (black circles), is plotted against the colored
noise thermostatted MTS prescription utilizing an outer time step of
12 fs (red solid line) and 20 fs (blue dashed line).  The inset
depicts the first peak of $g_\text{OH}(r)$ corresponding to the OH
covalent bond.} \label{fig:watergr}
\end{center}
\end{figure}

In Figure \ref{fig:watergr}, the set of radial distributions of water
of the GLE - 12 fs and GLE - 20 fs stabilized runs are shown to be in
excellent agreement with the baseline calculations.  The GLE - 16 fs
results for clarity are not shown but are fully consistent with the
other results.  The inset shows that the peak corresponding to the
oxygen hydrogen covalent bond is also well reproduced.  This finding
is significant since, as discussed above, the bonding energy tends to
show a larger error than the overall energy. This reflects the fact
that these components vary rapidly, thereby inducing a greater
sensitivity to small deviations in position.  Overall, the results of
Figure \ref{fig:watergr} underline the fact that the free energy
surface is accurately reproduced by the GLE resonance stabilized
simulations.

Although both white and colored noise may be utilized to stabilize
resonance artifacts, these two techniques significantly differ in the
degree to which they disturb dynamical properties. In Table
\ref{tab:tab3} we report the diffusion constant and the relaxation
time of the first order molecular dipole orientational correlation
function for the resonance stabilized simulations in comparison to the
microcanonical baseline results.  The colored noise stabilized
dynamics with an outer time step of 12 fs differ by $4\%$ from the NVE
results as compared to the difference of a factor of $2.5$ that is
obtained from the WNL - 12 fs run.  This finding is consistent with
the selectivity of a colored noise thermostat which only couples
weakly to the low frequency modes and therefore minimally perturbs
dynamical quantities that largely depend upon these slower modes (see
Fig. \ref{fig:spec}).  It can be seen from Table \ref{tab:tab3} that
the impact of the GLE - 12 fs profile on the dynamics is as good as
and in some cases better than that of the weak white noise friction
($\gamma = 1$ ps$^{-1}$) which is necessary to stabilize a simulation
that employs constraints at a 12 fs outer time step.~\cite{han07}
Therefore the present method of targeted damping of fast modes
compares very well to schemes where such modes are frozen.
 
The diffusion constant is extracted from the mean square displacement,
which aside from its physical meaning, is also an indicator of the
rate at which the simulation samples configuration space.  Therefore,
in addition to producing highly distorted dynamics, the strength of
white noise coupling significantly slows down the sampling rate,
largely counterbalancing any efficiency gained by utilizing a larger
outer time step.~\cite{izaguirre01} This drawback is alleviated by the
use of the colored noise thermostat, although it can be clearly seen
from Table \ref{tab:tab3} that the dynamical perturbation engendered
by our scheme increases with outer time step as a ``stronger'' colored
noise profile is necessary to stabilize the system.  This arises
naturally as a greater number of modes begin to resonate at larger
outer time steps, thereby requiring strong coupling to the bath across
lower frequency modes to damp out such artifacts.  In this manner,
increased disturbance of the dynamics is necessitated.  The dynamics
produced by the GLE - 16 fs and GLE - 20 fs runs differ from the
baseline results by approximately $20\%$ and $35\%$, respectively.  A
comparison of the ratios of the GLE to the microcanonical results in
Table \ref{tab:tab3} confirms that this degree of distortion can be
reasonably estimated from \EQ{eq:shift}.  However, if only equilibrium
properties are desired and the computational speed-up gained by
increasing the outer time step outweighs the decrease in the sampling
rate, these choices may still offer advantages.

\begin{table*}
\begin{center}
\begin{tabular}{||c|c|c|c|c||} \hline thermostat & outer time step
(fs) & T (K) & $V_\text{nb}/N$ (kcal/mol) & $ V_\text{b}/N$ (kcal/mol)
\\ \hline WNL & 2 & $300.0$ & $ -3.872 $ & $0.5227$ \\ GLE & 12 &
$301.5$ & $ -3.869 $ & $ 0.5245 $ \\ GLE & 16 & $301.4$ & $ -3.869 $ &
$ 0.5247 $ \\ GLE & 20 & $302.0 $ & $ -3.872 $ & $ 0.5236 $ \\ WNL &
12 & $301.7$ & $ -3.875 $ & $ 0.5278 $ \\ WNL & 16 & $301.7$ & $
-3.874 $ & $ 0.5285 $ \\ WNL & 20 & $302.3$& $ -3.875 $ & $0.5246 $ \\
\hline
\end{tabular}
\caption{Average temperatures and bonded and non-bonded potential
energies per atom of the flexible water system for different
combinations of thermostat and outer time step.  The figures are
reported to a precision within the error of the
calculation.} \label{tab:tab2}
\end{center}
\end{table*}

\begin{table}
\begin{center}
\begin{tabular}{||c|c|c|c||} \hline thermostat & outer time step (fs)
& D ($\text{\AA}^2$/ps) & $\tau_\text{dipole}$ (ps) \\ \hline NONE & 2
& $0.25 $ & $4.5 $ \\ WNL & 2 & $ 0.22 $ & $4.8 $ \\ GLE & 12 & $0.24
$ & $4.7 $ \\ GLE & 16 & 0.20 & 5.6 \\ GLE & 20 & 0.16 & 6.7 \\ WNL &
12 & 0.095 & 8.0 \\ WNL & 16 & 0.044 & 14 \\ WNL & 20 & 0.019 & 26 \\
\hline
\end{tabular}
\caption{The dynamic quantities computed in the flexible water system.
The diffusion constant and the first order molecular dipole relaxation
time are given above. Figures are reported to statistical
accuracy.} \label{tab:tab3}
\end{center}
\end{table}

\subsection{Alanine} \label{ssec:alan} Multiple time scale techniques
are often employed to simulate biomolecular systems.  In this section,
we gauge the applicability of our scheme to a fully flexible
simulation of alanine dipeptide in explicit water solvent.  The use of
this model system allows us to readily test the impact of the chosen
colored noise profiles on conformational sampling and dynamics.  The
colored noise profiles utilized are the same as those presented in
Section \ref{ssec:water} and here we show their applicability to more
general systems.

In Table \ref{tab:tab4} the energies of the alanine system for the
chosen combinations of thermostats and outer time steps are presented.
As in the case of pure water, the colored noise profiles reproduce the
average bonded and non-bonded potential energies of the baseline (WNL
- 2 fs) result to within $0.5\%$.  The bonded energies include the
stretches, bends, and torsions of alanine dipeptide in addition to the
intramolecular interactions of water. As also noted in Section
\ref{ssec:water}, the bonded energies, which contain high frequency
modes and most strongly exhibit the resonance phenomena, possess a
larger overall error than the non-bonded energies.  This behavior
underlines why our scheme of targeting fast motions for damping is
successful.  Additionally the total and alanine dipeptide temperature
are reported, and can also be seen to be within $0.5\%$ of the
baseline results.

\begin{figure}
\begin{center}
\includegraphics[scale=0.45]{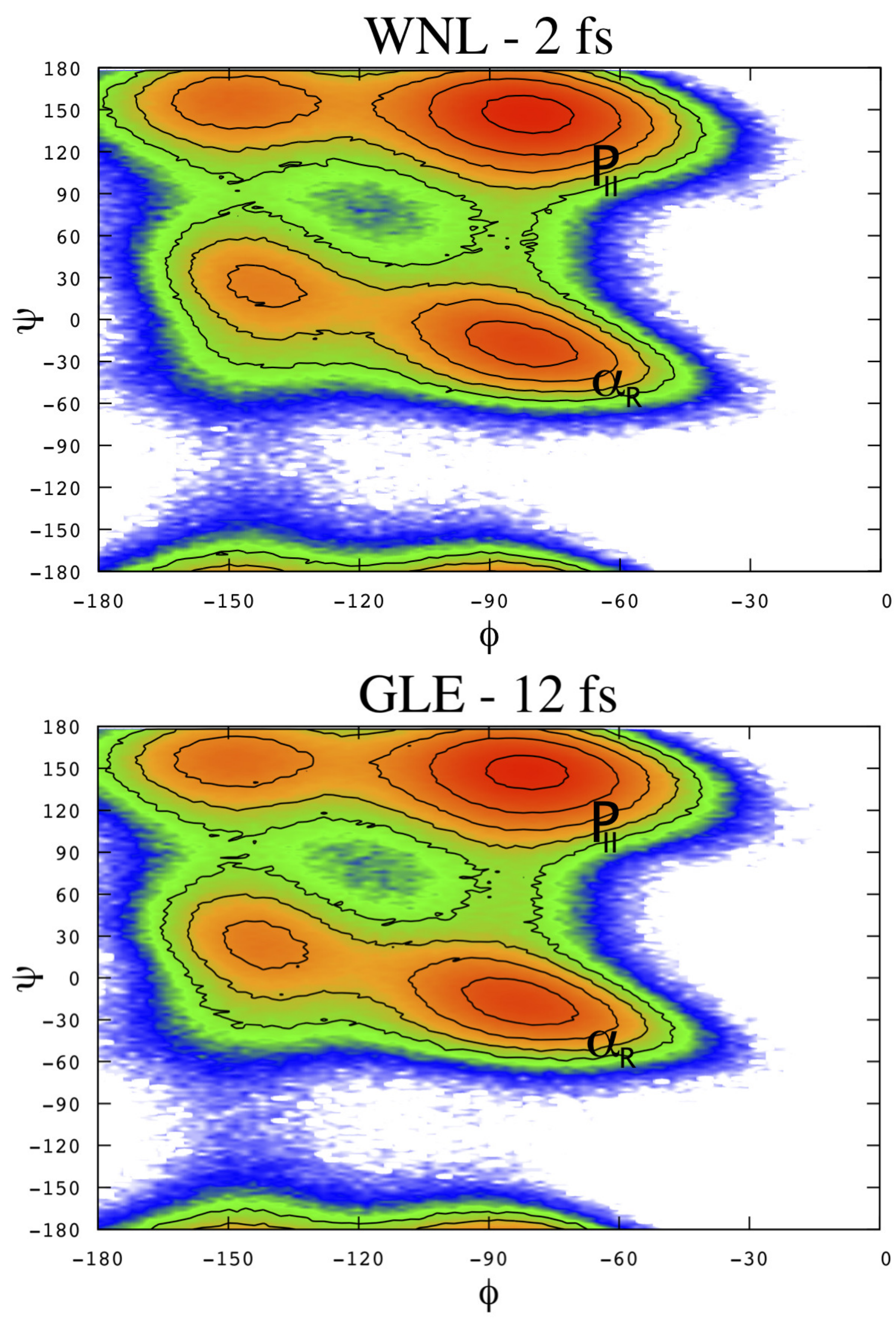}
\caption{The free energy as a function of dihedral angles $\phi$ and
$\psi$. The baseline WNL - 2fs (top panel) and GLE - 12fs (bottom
panel) results are shown. The free energy in conformation space
decreases as the color varies from white to red.  Isolines represent
increments of 0.5 kcal/mol.  The $\alpha_R$ and $\text{P}_\text{II}$ regions are
labeled in each panel.} \label{fig:rama}
\end{center}
\end{figure}

The conformational space of alanine dipeptide is typically
characterized as a function of the two dihedral angles $\psi$ and
$\phi$.~\cite{hummer03,chekmarev04,kwac08} The Ramachandran plots of the
baseline and GLE - 12 fs result are given in Figure \ref{fig:rama}.
It can be seen that they are in excellent agreement.  The resultant
free energy exhibits a minimum in both the $\alpha$-helical and the
extended $\text{P}_\text{II}$ region.  These two regions are labeled in Figure
\ref{fig:rama}.

The mean first passage time is estimated from the survival probability
of the transition between the $\alpha_R$ and the
$\text{P}_\text{II}$ regions.  The survival probability, $S(t)$, for a state in the
$\alpha_R$ region to cross to the $\text{P}_\text{II}$ region may be defined in terms
of an average in conformation space over trajectories that reside in
the $\alpha_R$ region at time $t=0$ and outside the $\text{P}_\text{II}$ region up to
time, $t$ and is given by,
\begin{eqnarray} S_{\alpha_R \rightarrow \text{P}_\text{II}}(t)&=& \left< 1 -
g_{\text{P}_\text{II}}(t) \right>_{h_{\alpha_R}(0)=1} ,
\end{eqnarray} 
where $h_{\alpha_R}$ is unity inside the $\alpha_R$ region and zero outside,
and $g_{\text{P}_\text{II}}(t)$ is defined to be unity if the particle has passed
into region $\text{P}_\text{II}$ at any time between 0 and $t$ and is zero
otherwise.  The $\psi$ range of the $\alpha$-helical region is defined
as $-40^\circ < \psi < 10^\circ$ and the extended $\text{P}_\text{II}$ region as
$110^\circ < \psi < 180^\circ$.  The $\phi$ range for both regions is
set to be $-110^\circ < \phi < -60^\circ $.

\begin{figure}
\begin{center}
\includegraphics[scale=0.35]{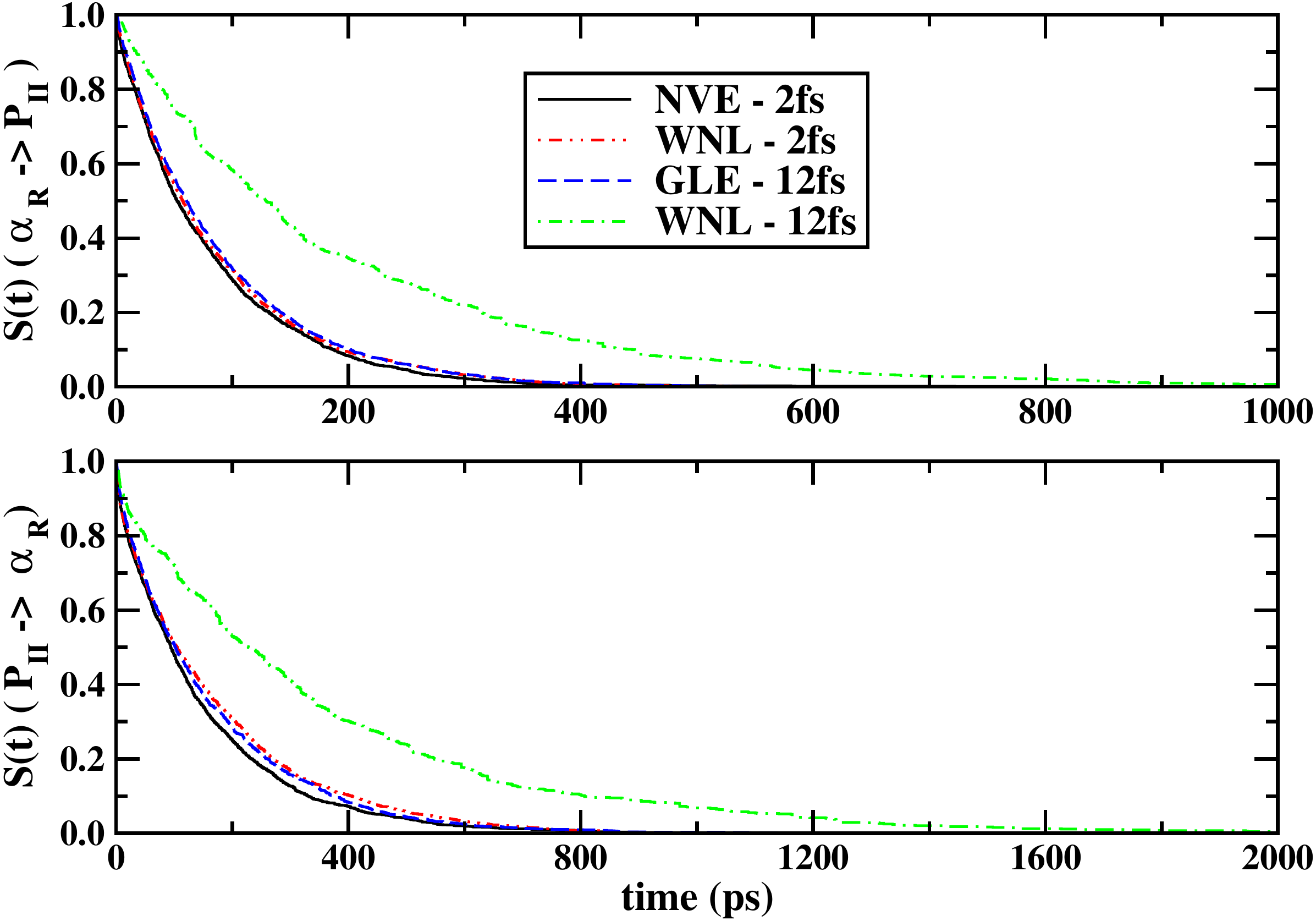}
\caption{The survival probabilities for selected runs of the alanine
dipeptide system are shown for the transition from the $\alpha_R$ to
$\text{P}_\text{II}$ region (top panel) and for the reverse process (bottom panel).
Results are given for the baseline microcanonical (solid black curve)
and white noise Langevin (red dot-dashed curve) runs as well as the
simulations that are resonance stabilized with an outer time step of
12 fs utilizing colored (blue dashed curve) and white (green
dot-dashed curve) noise.} \label{fig:surv}
\end{center}
\end{figure}

The survival probabilities for the $\alpha_R \rightarrow \text{P}_\text{II}$ and
$\text{P}_\text{II} \rightarrow \alpha_R$ transitions are plotted in Figure
\ref{fig:surv}.  The mean first passage times are presented in Table
\ref{tab:passage}.  The equilibrium constant of the $\alpha_R
\rightleftharpoons \text{P}_\text{II}$ process is related to the ratio of the
$\text{P}_\text{II} \rightarrow \alpha_R$ mean passage time to the $\alpha_R
\rightarrow \text{P}_\text{II}$ value and is $\approx 1.8-1.9$ in all runs.  Upon
study of the mean first passage times, it can be seen that the colored
noise thermostated result with an outer time step of 12 fs is
perturbed by $\approx 10\%$ with respect to the microcanonical result,
and is again comparable with the results of the system when weakly
coupled to a white noise bath ($\gamma = 1$ ps$^{-1}$).  The colored
noise thermostat causes a slightly larger perturbation on the mean
passage time when compared to that exhibited in the diffusion of water
(see Sec. \ref{ssec:water}).  This likely arises from the dependence
of this property on torsional motions that lie in a higher frequency
range than diffusive modes, and are therefore more strongly coupled to
the bath. The large white noise friction that is necessary to damp out
resonance instabilities at an outer time step of 12 fs again increases
passage times by a factor of $\approx 2.5$.  The impact of large WNL
couplings on the conformational dynamics has been noted in previous
work.~\cite{derre95} As in the case of pure water, the degree of
perturbation induced by the GLE thermostat increases as a greater
degree of energy stabilization is necessary at larger outer time
steps.  The similarity of the results obtained in this case and pure
water underlines the fact the thermostatting scheme presented in this
work is transferable between flexible water and typical solvated
biomolecules.

\begin{table*}
\begin{center}
\begin{tabular}{||c|c|c|c|c|c||} \hline thermostat & outer time step
(fs) & T(K) &T$_\text{ALA}$ (K) & $V_\text{nb}/N$ (kcal/mol) &
$V_\text{b}/N$ (kcal/mol) \\ \hline WNL & 2 & $300.0 $ & $299.7 $ & $
-3.828 $ & $ 0.5297 $ \\ GLE & 12 & $301.0 $ & $ 300.1 $ & $ -3.825 $
& $ 0.5312 $\\ GLE & 16 & $301.1$ & $299.8$ & $-3.823$ & $0.5314$ \\
GLE & 20 & $301.7$ & $299.8$ & $-3.826$ & $0.5304$ \\ WNL &12 & 301.3
& 300.0 & $-3.829$ & $0.5338$ \\ WNL & 16 & 301.4 & 299.8 & $-3.828$ &
$0.5344$ \\ WNL & 20 & $302.0 $ & $300.0$ & $-3.828$ & $0.5315$ \\
\hline
\end{tabular}
\caption{The total and alanine dipeptide temperatures, as well as the
bonded and non-bonded energy per atom are given for each combination
of outer time step and thermostatting scheme utilized for alanine
dipeptide in explicit solvent. The figures are reported to statistical
accuracy.} \label{tab:tab4}
\end{center}
\end{table*}

\begin{table*}
\begin{center}
\begin{tabular}{||c|c|c|c||} \hline thermostat & outer time step (fs)
& $\bar{\tau}(\alpha_R \rightarrow \text{P}_\text{II})$ (ps) & $\bar{\tau} (\text{P}_\text{II}
\rightarrow \alpha_R)$ (ps) \\ \hline NONE & 2 & 79 & 142 \\ WNL & 2 &
85 & 163 \\ GLE & 12 & 87 & 156 \\ GLE & 16 & 105 & 193 \\ GLE & 20 &
121 & 234 \\ WNL &12 & 190 & 340 \\ WNL & 16 & 435 & 790 \\ \hline
\end{tabular}
\caption{Table of mean first passage times, $\bar{\tau}$, from the
$\alpha_R$ to $\text{P}_\text{II}$ and $\text{P}_\text{II}$ to $\alpha_R$ regions are presented for
various combinations of outer time steps and thermostatting schemes.
The average error for runs where the dynamics are not strongly
overdamped is $\approx 3$ ps.  Results are not given for the WNL - 20
fs run due to the fact that too few transitions occurred in the course
of the simulation to generate a reliable estimate of
$\bar{\tau}$. } \label{tab:passage}
\end{center}
\end{table*}

\section{Conclusions} \label{sec:conc} In this paper, we have shown
that by coupling a standard multiple time scale scheme to a colored
noise bath large outer time steps can be employed while obtaining
accurate sampling and minimally perturbing the dynamics. The scheme
was illustrated with applications to flexible water and a fully
flexible simulation of aqueous alanine dipeptide. Our results suggest
that, when combined with our scheme, a 12 fs outer time step seems a
good compromise between size of outer time step and minimizing the
impact on the dynamics. With this outer time step the damping needed
to obtain potential energies within $0.5 \%$ of the benchmark values
decreased diffusion by only $4 \%$. In contrast, stabilization of the
simulation at this outer time step using white noise decreases the
diffusion constant by over 2.5 times. Even more promising, the
dynamical perturbation due to our colored noise scheme compared
favorably with the weak white noise damping ($\gamma=1$ ps$^{-1}$)
required to stabilize a multiple time scale simulation in which all
bonds to hydrogen are constrained.~\cite{han07} This suggests that, in
conjunction with multiple time scale algorithms, this method may be
utilized in lieu of constraints.

In our serial code using a 12 fs outer time step yielded a
computational speed-up of 4.3 times compared to our baseline MTS
calculations using a 2 fs outer time step and a 17 times speed up
compared to flexible simulations without a multiple time scale scheme
that employ a 0.5 fs time step. Even further increases may be achieved
when the computational architecture or parallelization scheme makes
long range electrostatics comparatively costly to
evaluate.~\cite{desmond} It is therefore clear that this scheme
provides an efficient approach to perform equilibration or
configurational sampling. Although our approach perturbs the dynamics
we note that the alanine dipeptide molecule used in our benchmark
simulation was specifically chosen to allow us to view many
transitions and hence converge the dynamical properties within small
bounds. In typical, large-scale biological system where long
time-scale processes are of interest this would not be the case and
hence the small change in the dynamics due to this method will likely
be dwarfed by the statistical errors. In this case the ability to
generate longer trajectories using this approach facilitates a
decrease in the statistical errors bars.

The similarities in the spectra of many aqueous and biological systems
suggest that our noise profiles should provide good out-of-the box
performance for other fully flexible systems. However, the flexibility
of the colored noise approach affords such a large degree of
versatility that further tuning may yield improved results. Our
matrices may also provide a basis for other applications where energy
must be rapidly dissipated to ensure accurate sampling such as in the
case of fast-deposition metadynamics.~\cite{ceriotti_thesis} The
methods outlined in Sec. II and Appendix \ref{app:forms} provide a
starting point for such developments.
 
\begin{acknowledgements} This research was supported from a grant to
B.J.B.  from the National Science Foundation via grant
NSF-CHE-0910943.
\end{acknowledgements}

\appendix
\section{Further applications of simple drift
matrices} \label{app:forms} It is often convenient to consider drift
matrices whose elements are transparently related to the time
dependence of the memory kernel (i.e. the eigenvalues of submatrix
$\bmat{\Gamma}_{ss}$).  This facilitates the use of readily tunable
colored noise profiles that are based on a small set of parameters.
In addition to the form presented in this work (\EQ{eq:3x3}), two
other of such matrices have appeared in the literature.  The drift
matrix for a simple exponential noise is given by~\cite{ceriotti_09_1}
\begin{eqnarray} \bmat{\Gamma}_\text{A} &=& \left( \begin{array}{cc} 0
& \sqrt{\gamma_0 a} \\ -\sqrt{ \gamma_0 a} & a \end{array}
\right) \label{eq:gamma_a}
\end{eqnarray} where the memory function and spectra are:
\begin{eqnarray} K_\text{A}(t) &=& \gamma_0 a e^{-|t| a} \\
\hat{K}_\text{A}(\omega) &=& 2 \gamma_0 a \frac{a}{a^2 + \omega^2}.
\end{eqnarray} Whereas the following drift
matrix,~\cite{ceriotti_10_2}
\begin{widetext}
\begin{eqnarray} \bmat{\Gamma}_\text{B} &=& \left( \begin{array}{ccc}
0 & \sqrt{\frac{\gamma_0 (a^2+b^2)}{2a} } & \sqrt{\frac{\gamma_0
(a^2+b^2)}{2a}} \\ -\sqrt{\frac{\gamma_0 (a^2+b^2)}{2a}} & a & b \\
-\sqrt{\frac{\gamma_0 (a^2+b^2)}{2a}} & -b & a \end{array}
\right), \label{eq:gamma_b}
\end{eqnarray} 
corresponds to a term of damped oscillatory noise whose
spectra is centered at $\pm b$:
\begin{eqnarray} K_\text{B}(t) &=& \gamma_0 \frac{a^2+b^2}{a}
e^{-|t|a} \cos(b t) \\ \hat{K}_\text{B}(\omega) &=& \gamma_0
\frac{a^2+b^2}{a} \left( \frac{a}{a^2+\left(\omega - b \right)^2} +
\frac{a}{a^2 + \left(\omega + b \right)^2} \right). \label{eq:spec_b}
\end{eqnarray} 
\end{widetext}
In the above equations, the parameter, $a$, corresponds
to the real part and $b$ to the magnitude of the imaginary part of the
eigenvalue(s) of submatrix, $\bmat{\Gamma}_{ss}$.  As in Section
\ref{sec:theory}, the parameter $\gamma_0$ is proportional to the
value of the memory spectra at $\omega=0$, such that
$\hat{K}(0)=2\gamma_0$.

Drift matrices that correspond to stable dynamics may be combined such
that the resultant memory function is a weighted sum of the
corresponding memory functions of each component.~\cite{ceriotti_10_2}
For example, the drift matrix that corresponds to the sum of the noise
profiles of whose drift matrices are given by $^1\bmat{\Gamma}$ and
$^2\bmat{\Gamma}$ is given by,
\begin{eqnarray} ^{(1+2)}\bmat{\Gamma} &=& \left( \begin{array}{ccc}
(^1\gamma_{pp} + ^2\gamma_{pp}) & ^1 \bm{\gamma}_{ps}^T & ^2
\bm{\gamma}_{ps}^T \\ ^1\bm{\gamma}_{sp} & ^1\bmat{\Gamma}_{ss} &
\bmat{0} \\ ^2\bm{\gamma}_{sp} & \bmat{0} &
^2\bmat{\Gamma}_{ss} \end{array} \right) \label{eq:comb}
\end{eqnarray} where components are expressed in the notation of
\EQ{eq:acomp} .  In this way, forms such as those presented here may
be utilized as ``building blocks'' for more flexible colored noise
profiles.

Making use of this machinery, it is possible to design a form that
couples strongly to certain modes and weakly to others.  \EQ{eq:comb}
may be applied to add together noise profiles of the damped
oscillatory (\EQ{eq:spec_b}) or the ``canceling'' (\EQ{eq:3x3spec})
type. It must be noted that this formalism has drawbacks when compared
to the profiles utilized in the main text.  Namely, the dimensionality
of the resultant drift matrix grows with the number of targeted modes
and that the greater sensitivity to the details of the spectra implies
less robust performance across different systems.  However, this
sensitivity also facilitates fine control of the colored noise profile
for very specific applications.

\section{Dynamical properties of a GLE dynamics} \label{app:shift} In
order to estimate the perturbation on the dynamical properties of a
system caused by coupling to a colored noise thermostat one can study
the behavior of a one-dimensional harmonic model. In this limit,
microcanonical dynamical results in a $\delta$-function power spectrum
peaked at the oscillator's characteristic frequency $\omega_{0}$. The
presence of the noise will modify the line shape of the peak, shifting
its center and broadening it. A measure of these two effects can be
used to estimate the magnitude of the disturbance.

To achieve this, one must consider the matrices
$\boldsymbol{\Gamma}_{qp}\left(\omega_{0}\right)$ and
$\mathbf{B}_{qp}\left(\omega_{0}\right)$ which describe the dynamics
in the full $\mathbf{x}=\left(q,p,\mathbf{s}\right)$
space.~\cite{ceriotti_10_1} One can then find the stationary
covariance matrix $\mathbf{C}_{qp}$ by solving
$\boldsymbol{\Gamma}_{qp}\mathbf{C}_{qp}+\mathbf{C}_{qp}\boldsymbol{\Gamma}_{qp}^{T}=\mathbf{B}_{qp}\mathbf{B}_{qp}^{T}$
and compute the first order correlation matrix $\left\langle
\mathbf{x}^{T}\left(t\right)\mathbf{x}\left(0\right)\right\rangle $
and its Fourier transform,
\begin{eqnarray}
\mathcal{C}_{ij}\left(\omega\right)=\left[\frac{\boldsymbol{\Gamma}_{qp}}{\boldsymbol{\Gamma}_{qp}^{2}+\omega^{2}}\mathbf{C}_{qp}\right]_{ij}\left[\left(\mathbf{C}_{qp}\right)_{ii}\left(\mathbf{C}_{qp}\right)_{jj}\right]^{-1/2}.
\end{eqnarray}
The position and value at the maximum of the
$\mathcal{C}_{pp}\left(\omega\right)$ term can then be used to
characterize the deformed peak.

When $\boldsymbol{\Gamma}_{qp}$ is built out of the $3\times3$
canceling noise matrix in \EQ{eq:3x3} and we set $\gamma_{0}=0$ so as
to consider the case of small disturbance on the low-frequency modes
we obtain,
\begin{eqnarray}
\boldsymbol{\Gamma}_{qp}=\left(\begin{array}{cccc} 0 & -1& 0 & 0\\
\omega_{0}^{2} & \gamma & \sqrt[4]{3}\sqrt{\gamma\tilde{\omega}} &
\sqrt{\gamma\tilde{\omega}}/\sqrt[4]{3}\\ 0 &
\sqrt[4]{3}\sqrt{\gamma\tilde{\omega}} & \sqrt{3}\tilde{\omega} &
\tilde{\omega}\\ 0 & -\sqrt{\gamma\tilde{\omega}}/\sqrt[4]{3} &
-\tilde{\omega} & 0\end{array}\right).
\end{eqnarray}
Solving for
$\mathcal{C}_{pp}\left(\omega\right)$ we obtain
\begin{widetext}
\begin{eqnarray}
\mathcal{C}_{pp}\left(\omega\right)&=& \frac{3\gamma\omega^{6}}{\gamma^{2}\omega^{4}\left(3\omega^{2}+\tilde{\omega}^{2}\right)+2\sqrt{3}\gamma\tilde{\omega}\omega^{2}\left(\omega^{2}-\omega_{0}^{2}\right)\left(2\omega^{2}+\tilde{\omega}^{2}\right)+3\left(\omega^{2}-\omega_{0}^{2}\right)^{2}\left(\omega^{4}+\tilde{\omega}^{2}\omega^{2}+\tilde{\omega}^{4}\right)}
\end{eqnarray}
\end{widetext}

We then take $\tilde{\omega}$ to be much larger than all other
frequencies entering our problem, and write them as a ratio with
respect to $\tilde{\omega}$, i.e.
$\omega_{0}\leftarrow\omega_{0}/\tilde{\omega}$,
$\gamma\leftarrow\gamma/\tilde{\omega}$,
$\omega\leftarrow\omega/\tilde{\omega}$. One then finds the relevant
extremal point, $\omega_{P}$, and an estimate of the peak width as
$\Delta_{P}=1/\left[\pi\mathcal{C}_{pp}\left(\omega_{P}\right)\right]$.
This leads to expressions which can then be expanded in powers of
$\omega_{0}$, eventually yielding
\begin{eqnarray}
\omega_{P}=\frac{\omega_{0}}{\sqrt{1+\frac{\gamma/\tilde{\omega}}{\sqrt{3}}}}+\mathcal{O}\left(\left(\omega_0/\tilde{\omega}\right)^{2}\right)
\end{eqnarray}
and
\begin{eqnarray}
\Delta_{P}=\frac{\gamma\left(\omega_{0}/\tilde{\omega}\right)^{4}}{\pi\left(1+\frac{\gamma/\tilde{\omega}}{\sqrt{3}}\right)^{2}}+\mathcal{O}\left(\left(\omega_0/\tilde{\omega}\right)^{6}\right). \\[0.1in]
\nonumber
\end{eqnarray}

%bibliography

\newpage
\end{document}